\documentclass[english,nofootinbib,notitlepage]{revtex4}
\usepackage[T1]{fontenc}
\usepackage[latin1]{inputenc}
\usepackage{amsmath}
\usepackage{amssymb}
\usepackage{babel}
\usepackage{array}
\usepackage{ifpdf}
\ifpdf   
    \usepackage[pdftex]{graphicx}
    \usepackage[hyperfootnotes=false,pdfstartview=,colorlinks=true,linkcolor=purple,citecolor=blue,urlcolor=blue,bookmarks=false]{hyperref}    \DeclareGraphicsExtensions{.png,.jpg,.pdf}
    \usepackage{pgf}
\else    
    \usepackage[usenames,dvipsnames]{color}
    \usepackage[hyperfootnotes=false,pdfstartview=,colorlinks=true,linkcolor=BrickRed,citecolor=blue,urlcolor=blue,bookmarks=false]{hyperref}
    \usepackage{graphicx}
    \DeclareGraphicsExtensions{.eps}   
\fi
\graphicspath{{figures/}}

\begin{document}

\setlength{\marginparwidth} { 2.5  cm}
\setlength{\marginparsep}   { 0.0  cm}
\setlength{\topmargin}      { -0.8  cm} 
\setlength{\parskip}        { 0.0  cm}
\setlength{\arrayrulewidth} { 0.5  pt}
\setlength{\doublerulesep}  { 0.0  pt}
\addtolength{\headheight}   { 5.0  pt}
\setlength{\textheight}     {23.0  cm} 
\setlength{\textwidth}      {16.0  cm} 

\title{Growth factor and galaxy bias from future redshift surveys: a study
on parametrizations}

\author{Cinzia Di Porto$^{1,2}$, Luca Amendola$^{2}$, Enzo Branchini$^{1}$}


\affiliation{$^{1}$Dipartimento di Fisica {}``E. Amaldi'', Universit\`a degli
Studi {}``Roma Tre'', via della Vasca Navale 84, 00146, Roma, Italy}

\affiliation{$^{2}$Institut f$\ddot{\textrm{u}}$r Theoretische Physik, Universit$\ddot{\textrm{a}}$t Heidelberg, Philosophenweg 16, 69120 Heidelberg, Germany and INAF/Roma}

\begin{abstract}
Many experiments in the near future will test dark energy through
its effects on the linear growth of matter perturbations. In this
paper we discuss the constraints that future large-scale redshift
surveys  can put on three different parameterizations of
the linear growth factor and how these constraints will help ruling
out different classes of dark energy and modified gravity models. We show that a scale-independent bias can be estimated
to a few percent per redshift slice by combining redshift distortions
with power spectrum amplitude, without the need of an external estimation.
We find that the growth rate can be constrained to within 2-4\% for
each $\Delta z=0.2$ redshift slice, while the equation of state $w$ and
the index $\gamma$ can be simultaneously estimated both to within 0.02. We also find that a constant dimensionless coupling between dark energy and dark matter
can be constrained to be smaller than 0.14. 
\end{abstract}

\maketitle

\date{\today{}}

\section{Introduction}

The linear growth rate of matter perturbations is one of the most
interesting observable quantities  since it allows to explore
the dynamical features related to the build-up of cosmic structures
beyond the background expansion.
For example it can be used to discriminate between cosmological models
based on Einstein's gravity  and alternative models like $f(R)$ modifications of gravity (see e.g.~\cite{defelice10}) or multi-dimensional scenarios like in the Dvali-Gabadaze-Porrati (DGP)~\cite{dvali00} theory (e.g.~\cite{wang08}
and references therein).
In addition, the growth rate is sensitive to dark energy clustering or
to dark energy-dark matter interaction. For instance,
in models with scalar-tensor couplings or in $f(R)$ theories the growth rate at early epochs can be larger than in $\Lambda$CDM models and can acquire a scale dependence \cite{diporto08,tsujikawa09,gannouji09} (see for instance \cite{amendola_book} for a review on dark energy).

Simultaneous information on geometry and growth rate can be
obtained by measuring the galaxy power spectrum or the 2-point correlation function and their anisotropies observed in redshift space. These redshift distortions arise from peculiar velocities that contribute,
together with the recession velocities, to the observed redshift. The net
effect is to induce a radial anisotropy in galaxy clustering that can be
measured from standard two-point statistics like the power spectrum or the
correlation function~\cite{hamilton98}. The amplitude of the anisotropy is
determined by the typical amplitude of peculiar velocities which, in linear theory, is set by the growth rate of perturbations\footnote{In order to avoid confusion with the $f(R)$ models, we use the letter $s$, for slope, rather than the more popular $f$, to indicate the growth rate.}:
\begin{equation}
s\equiv\frac{d\log G}{d\log a}\,,\end{equation}
where $G(z)\equiv\delta(z)/\delta(0)$ is the growth function,
$\delta(z)$ the matter density contrast and the scale factor $a$ is related to the redshift $z$ through $a=(1+z)^{-1}$. Since however we only observe the clustering of galaxies and not that of the matter, the quantity that is accessible to observations is actually
\begin{equation}
\beta\equiv\frac{s}{b}\,,\label{eq:beta_sb}
\end{equation}
 where the bias $b$ is the ratio of density fluctuations in galaxies and
matter. The bias is in general a function of redshift and scale, but in the following we will consider it as a simple scale-independent function.

Once the power spectrum is computed in $k$-space, the analysis proposed
in~\cite{seo03} can be exploited to constrain not only geometry
but also the growth rate (as pointed out in~\cite{amendola05}; see
also~\cite{sapone07,wang08}), provided
that the power spectrum is not marginalized over its amplitude.
In configuration space, the first analysis of the two-point correlation
function explicitly aimed at
discriminating models of modified gravity from the standard $\Lambda$CDM
scenario has been performed by~\cite{guzzo08}.
Currently, there are several  experimental estimates of
the growth factor derived from the analysis of the redshift
space distortions~\cite{hawkins03,verde02,tegmark06,ross07,guzzo08,daangela08,peacock01,blake10},
from the redshift evolution of the  {\it rms} mass fluctuation
$\sigma_8$ inferred from
Ly$\alpha$ absorbers~\cite{mcdonald05} and from the power spectrum of density 
fluctuations measured from galaxies' peculiar velocities~\cite{nusser11}.
Current uncertainties are still too large to allow these measurements to
discriminate among alternative cosmological scenarios.
(e.g.~\cite{nesseris08,dossett10}).

On-going redshift surveys like VIPERS~\cite{guzzo10} or BOSS~\cite{BOSS}
will certainly provide more stringent constraint and will be able to test those models that deviate most from the standard cosmological model. However, only next generation large-scale redshift surveys at $z\approx1$ and beyond like EUCLID~\cite{euclid} or BigBOSS~\cite{BigBOSS} will provide an efficient way to discriminate competing dark energy models.

The growth rate $s$ clearly depends on the cosmological model. It
has been found in several works
\cite{peebles76,lahav91,polarski08,linder05,wang98}
that a simple yet effective parameterization of $s$ captures the behavior of
a large class of models. Putting
\begin{equation}
s=\Omega_{m}^{\gamma}\,,\label{eq:standard}
\end{equation}
where $\Omega_{m}(z)$ is the matter density in units of the critical
density as a function of redshift, a value $\gamma\approx0.545$ reproduces
well the $\Lambda$CDM behavior while departures from this value characterize
different models. For instance the DGP  is well approximated by
$\gamma\approx0.68$~\cite{linder07,wei08}
while viable models of $f(R)$ are approximated by $\gamma\approx0.4$
for small scales and small redshifts~\cite{tsujikawa09,gannouji09}.
This simple parameterization is however not flexible enough to accommodate
all cases. A constant $\gamma$ cannot for instance reproduce a
growth rate larger than $s=1$ in the past (as we have in $f(R)$ and
scalar-tensor
models) allowing at the same time $s<1$ at the present epoch if
$\Omega_{m}\le1$. Even in standard cases, a better approximation
requires a slowly-varying, but not strictly constant, $\gamma$.

In addition, the measures of the growth factor obtained from redshift
distortions require an estimate of the galaxy bias, which can be obtained either independently, using higher order statistics (e.g.~\cite{verde02,marinoni05}) or inversion techniques~\cite{sigad00}, or self consistently, by assuming some reasonable form for the bias function {\it a priori} (for instance, that the bias is independent of scale, as we will assume here).

The goal of  this paper is to forecast the constraints that
future observations can put on the growth rate. In particular we use
representative
assumptions for the parameters of the EUCLID survey to provide a baseline
for future experiments
and we focus on the following issues. {\it i}) We assess how well one can
constrain the bias function
from the analysis of the power spectrum itself and evaluate the impact that
treating bias as a free
parameter has on the estimates of the growth factor. We compare the results
with those obtained under the more popular approach 
of fixing the bias factor (and its error) to some independently-determined value.
{\it ii}) We estimate how errors depend on the parameterization of the growth
factor and on the
number and type of degrees of freedom in the analysis. {\it iii}) We
explicitly explore the case of
coupling between dark energy and dark matter and assess the ability of
measuring the coupling constant.

We do this in the context of the Fisher Matrix analysis. This is a common
approach that has been adopted
in several recent works, some of which exploring the case of a EUCLID-like
survey as we do.
We want to stress here that this work is, in fact, complementary to those of
the other authors.
Unlike most of these works, here we do not try
to optimize the parameter of the EUCLID survey in order to improve the
constraints on the
relevant parameters, as in~\cite{wang10}. Instead, we adopt a
representative sets of parameters
that describe the survey and derive the expected errors on the interesting
quantities.
In addition, unlike \cite{simpson10}
and ~\cite{samushia10}, we do not explicitly aim to study the correlation
between the parameters that describe the geometry of the system and
the growth parameters, although in our approach we also take into
account the degeneracy between geometry and growth.
Finally, the main results of this paper are largely complementary to the work of~\cite{majerotto11} that perform a more systematic error analysis that does not cover the main issues of our work.

Although, as we mentioned, in general $s$ might depend on scale,
we limit this paper to an exploration of time-dependent functions
only. Forecasts for specific forms of scale-dependent growth factor
motivated by scalar-tensor models are in progress and will be presented
elsewhere.

The layout of the paper is as follows. In the next section we will introduce the different parameterizations adopted for the growth rate and for the equation of state of dark energy, together with the models assumed for the biasing function, and describe the different cosmological models we aim to discriminate. In sec.~\ref{sec:fm} we will briefly review the Fisher matrix method for the power spectrum and define the adopted fiducial model. In sec.~\ref{sec:survey} we will describe the characteristics of the galaxy surveys considered in this work, while in sec.~\ref{sec:results} we will report our results on the forecast errors on the parameters of interest. Finally, in sec.~\ref{sec:conclusions} we will draw our conclusions and discuss the results.

\section{Models}\label{sec:models}

The main scope of this work is to quantify the ability of future 
redshift surveys to constrain the growth rate of density fluctuations. In particular we want to quantify how this ability depends on the parameterization assumed for $s$ and for the equation of state of the dark energy $w$ and on the biasing parameter. For this reason we explore different scenarios detailed below.

\subsection{Equation of state}

\begin{itemize}
\item{\it $w$-parameterization}. In order to represent the evolution of the equation of state parameter $w$, we use the popular CPL parameterization~\cite{chevallier01,linder03} 
\begin{equation}
w(z)=w_{0}+w_{1}\frac{z}{1+z}\;.\label{eq:w_CPL}
\end{equation}
As a special case we will also consider the case of a constant $w$.
\end{itemize}

\subsection{Growth Rate}\label{subsec:parameterizations}
As anticipated, in this work we assume that the growth rate, $s$, is a function of time
but not of scale. Here we explore three different parameterizations of $s$:

\begin{itemize}

\item {\it $s$-parameterization}. This is in fact a non-parametric 
model in which the growth rate itself is modeled as a step-wise function $s(z)=s_{i}$, specified in different redshift bins. The errors are derived on $s_{i}$ in each $i$-th redshift bin of the survey. 

\item {\bf {\it $\gamma$-parameterization}}.
As a second case we assume
\begin{equation}
s\equiv\Omega_{m}(z)^{\gamma(z)}\;.\label{eq:s_parametriz}
\end{equation}
where the $\gamma(z)$ function is parametrized as 
\begin{equation}
\gamma(z)=\gamma_{0}+\gamma_{1}\frac{z}{1+z}\;.\label{eq:gam_CPL}
\end{equation}
As shown by \cite{wu09,fu09}, this parameterization is more accurate than that of eq.~(\ref{eq:standard}) for both $\Lambda$CDM and DGP models.
Furthermore, this parameterization is especially effective to distinguish between a $w$CDM model  (i.e. a dark energy model with a constant equation of state) that has a negative $\gamma_{1}$ 
($-0.020\lesssim\gamma_{1}\lesssim-0.016$, for a present matter density $0.20\leq\Omega_{m,0}\leq 0.35$) and a DGP model that instead, has 
a positive $\gamma_{1}$ ($0.035<\gamma_{1}<0.042$).
In addition, modified gravity models show a strongly evolving $\gamma(z)$~\cite{gannouji09,motohashi10,fu09}, in contrast with conventional Dark Energy models.
As a special case we also consider $\gamma=$ constant (only when $w$ also is assumed constant), to compare our results with those of previous works.

\item {\bf {\it $\eta$-parameterization}}.
To explore models in which perturbations grow faster than in the 
$\Lambda$CDM case, like in the case of a coupling between dark energy and dark matter~\cite{diporto08}, we consider a model in which $\gamma$ is constant and the growth rate varies as
\begin{equation}
s\equiv\Omega_{m}(z)^{\gamma}(1+\eta)\;,\label{eq:eta_paramet}
\end{equation}
where $\eta$ quantifies the strength of the coupling.
The example of the coupled quintessence model worked out by~\cite{diporto08} illustrates this point. In that model, the numerical solution for the growth rate can be fitted by the formula (\ref{eq:eta_paramet}), with 
$\eta=c\beta_{c}^{2}$,
where $\beta_{c}$ is the dark energy-dark matter coupling constant
and  best fit values  $\gamma=0.56$ and $c=2.1$.
In this simple case, observational constraints over
$\eta$ can be readily  transformed into constraints over $\beta_{c}$.

\end{itemize}

\subsection{Galaxy Biasing}

In the analysis of the redshift distortions, $s(z)$ is degenerate with the bias function $b(z)$.
In absence of a well-established theory of galaxy formation and
evolution, most analysis assume some arbitrary functional form for $b(z)$. 
However, biasing needs to be neither deterministic nor linear.
Stochasticity in galaxy biasing is supposed to have little impact
on two-point statistics, at least on scales significantly larger than
those involved with galaxy evolution processes~\cite{dekel99}.
On the other hand, deviations from linearity (which imply scale dependency) might not be negligible. Current observational constraints 
based on self consistent biasing estimators~\cite{sigad00,verde02}
show that nonlinear effects are of the order of a few to $\sim 10$ \%, depending on the scale and the galaxy type~\cite{marinoni05,kovac09}.
To account for current uncertainties in both modeling and measuring 
galaxy bias we consider the following choices for the functional form of $b$:

\begin{itemize}
\item{\it{Redshift dependent bias}}.
We assume  $b(z)=\sqrt{1+z}$ (already used in~\cite{rassat08}) since this function provides a good fit to H$_{\alpha}$ line galaxies with luminosity $L_{{\rm H}_{\alpha}}=10^{42}$ erg$^{-1}$ s$^{-1}$ h$^{-2}$ modeled by~\cite{orsi10} using  the semi-analytic
$GALFORM$ models of~\cite{baugh05}.
We consider H$_{\alpha}$ line objects since they will likely constitute
the bulk of galaxies in the next generation slitless spectroscopic surveys like EUCLID. This  
H$_{\alpha}$ luminosity roughly 
corresponding, at $z=1.5$,
to a limiting flux of $f_{{\rm H} \alpha} \ge 4\times10^{-16}$  $ \textrm{erg } \textrm{cm}^{-2}  \textrm{s}^{-1}$.

\item{\it{Constant bias}}.
For the sake of comparison, we will also consider the case of constant
$b=1$ corresponding to the rather unphysical case of a redshift-independent population of unbiased mass tracers.
\end{itemize}

\subsection{Reference Cosmological Models}
As it will be better explained in the next section, to perform the Fisher Matrix analysis we need to adopt a fiducial cosmological model. We choose the one
recommended by the Dark Energy Task Force (DETF)~\cite{DETF}.
In this ``pseudo'' $\Lambda$CDM model the growth rate values are obtained from
eq.~(\ref{eq:standard}) with $\gamma=0.545$ and $\Omega_m(z)$ is given by the standard evolution
\begin{equation}
\Omega_{m}(z)=\Omega_{m,0}(1+z)^3\frac{H_0^2}{H(z)^2}\,,\label{eq:omega_standard}
\end{equation}
where (the subscript $0$ will generally denotes the present value)
\begin{equation}
H(z)^2=H_0^2\left[\Omega_{m,0}(1+z)^3+\Omega_{k}(1+z)^2 +(1-\Omega_{m,0}-\Omega_{k})\exp\left\{3\int\left(1+w_{0}+w_{1}\frac{z}{1+z}\right)\frac{dz}{1+z}\right\}\right]\label{eq:H_std}\;.
\end{equation}
Then $\Omega_m(z)$ is completely specified by setting $\Omega_{m,0}=0.25$,  $\Omega_k=0$, $w_0=-0.95$, $w_1=0$.
We wish to stress that regardless of the parameterization adopted, our fiducial cosmology is always chosen as the DETF one. In particular we choose as fiducial
values $\gamma_{1}=0$ and $\eta=0$, when the corresponding parameterizations are employed.

One of the goals of this work is to assess whether the
analysis of the power spectrum in redshift-space
can distinguish the fiducial model from alternative
cosmologies, characterized by their own set of parameters
(apart from $\Omega_{m,0}$ which is set equal to 0.25 for all of them). The
alternative models that we consider in this work are:

\begin{itemize}

\item {\it DGP model}. We consider the flat space case studied in~\cite{maartens06}.
When we adopt this model then we set $\gamma_{0}=0.663$,\,$\gamma_{1}=0.041$~\cite{fu09} or $\gamma=0.68$~\cite{linder07} and $w=-0.8$ when $\gamma$ and $w$ are assumed constant.

\item {\it $f(R)$ model}. Here we consider the one proposed in 
\cite{hu07}, depending on two parameters, $n$ and $\lambda$, which we fix to $n=2$ and $\lambda=3$. In this case we assume $\gamma_{0}=0.43$, $\gamma_{1}=-0.2$, values that apply quite generally in the limit of small scales (provided they are still linear, see \cite{gannouji09}) or $\gamma=0.4$ and $w=-0.99$.

\item {\it coupled dark energy (CDE) model}.  This is the coupled model 
proposed by \cite{amendola00,wetterich95}. In this case we assume $\gamma_{0}=0.56$, $\eta=0.056$ (this value comes from putting $\beta_{c}=0.16$ as coupling, which is of the order of the maximal value allowed by CMB constraints)~\cite{amendola03}. As already explained, this model cannot be reproduced by a constant $\gamma$.

\end{itemize}

\section{Fisher Matrix Analysis}\label{sec:fm}

In order to constrain the parameters, we use the Fisher matrix method~\cite{fisher35} (see~\cite{tegmark97r} for a review), that we apply to the power spectrum analysis in redshift space following~\cite{seo03}.
For this purpose we need an analytic model of the power spectrum
in redshift space as a function of the parameters that we wish to constrain.
The analytic model is obtained in three steps. ({\it i} ) First of all we 
compute with CMBFAST~\cite{seljak96} the linear power spectrum of the matter in real space at $z=0$, $P_{0r}(k)$, choosing a reference cosmology where the parameters to be given as input (i.e. $\Omega_{m,0} h^2$, $\Omega_{b,0} h^2$, $h$, $n_s$ also employed in the Fisher matrix analysis, plus the other standard parameters
required by the CMBFAST code) are set to the values given in the III column of Tab.~\ref{tab:parameters} while for the normalization of the spectrum we use $\sigma_8 =0.8$. ({\it ii} ) Second, we model the linear redshift-space distortions as 
\begin{equation}
P_{\textrm{obs}}(z;k,\mu)=P_{0r}(k)\frac{D_{F}^{2}(z)H(z)}{D^{2}(z)H_{F}(z)}G^{2}(z)b^{2}(z)(1+\beta(z)\mu^{2})^{2}+P_{\textrm{s}}(z)\;.\label{eq:P_obs}
\end{equation}
Here the subscript $F$ indicates quantities evaluated on the fiducial
model. 
In this expression $H(z)$ is the expansion history in Eq.~\ref{eq:H_std}, 
$D(z)$ is the angular diameter distance, $G(z)$ the growth factor and $P_{s}(z)$ represents a scale-independent offset due to imperfect removal of shot-noise.
Finally $\beta(z)$ is the redshift distortion parameter and the term
$(1+\beta\mu^{2})^{2}$ is the factor invoked by \cite{kaiser87}
to account for linear distortion in the distant-observer's approximation, where $\mu$ is the  direction cosine of the wavenumber ${\boldsymbol k}$ with respect to the line of sight.
As shown in~\cite{amendola05,sapone07} and recently in~\cite{wang10}, the 
inclusion of growth rate information reduces substantially the errors on the parameters, improving the figure of merits.
({\it iii} ) As a third and final step we account for nonlinear effects.
On scales larger than ($\sim 100\, h^{-1}$Mpc) where we focus our analysis, nonlinear effects can be represented as a displacement field in Lagrangian space modeled by an elliptical Gaussian function. Therefore, following \cite{eisenstein07,seo07}, to model nonlinear effect
we multiply $P_{0r}(k)$ by the factor
\begin{equation}
\exp\left\{ 
-k^{2}\left[\frac{(1-\mu^{2})\Sigma_{\perp}^{\,2}}{2}+\frac{\mu^{2}\Sigma_{\parallel}^{\,2}}{2}\right]\right\}
\,, \label{eq:damping}
\end{equation}
where $\Sigma_{\perp}$ and $\Sigma_{\parallel}$ represent the
displacement across and along the line of sight, respectively. They  are related to the growth factor $G$ and to the growth rate
$s$ through $\Sigma_{\perp}=\Sigma_{0}G$ and $\Sigma_{\parallel}=\Sigma_{0}G(1+s)$.
The value of $\Sigma_{0}$ is proportional to $\sigma_{8}$. For our 
reference cosmology where $\sigma_{8}=0.8$~\cite{wmap7}, we have $\Sigma_{0}=11\, h^{-1}$Mpc.

The observed power spectrum in a given redshift bin depends therefore
on a number of parameters, denoted collectively as $p_{i}$, such as the Hubble constant at present $h$, the reduced matter and baryon fractions at present, $\Omega_{m,0}h^2$ and $\Omega_{b,0}h^2$, the curvature density parameter $\Omega_{k}$, the spectral tilt $n_{s}$ plus the parameters that enter in the parameterizations described in the previous section: $w_0$, $w_1$ (or simply $w$); $\gamma_0$, $\gamma_1$ (or $\gamma$) and  $\eta$. They are listed in Tab.~\ref{tab:parameters} and are referred to as ``Cosmological parameters''. These parameters will be left free to vary while we always fix $\sigma_8$=0.8 since the overall amplitude is degenerate
with growth rate and bias. 
The other free parameters depend on the redshift. They are listed in the lower part of Tab.~\ref{tab:parameters} and include
the expansion history $H(z)$, the growth factor $G(z)$, the angular diameter distance $D(z)$, the shot noise $P_{s}(z)$, the growth rate $s(z),$ the redshift distortion parameter $\beta(z)$ and the galaxy bias $b(z)$.

Given the model power spectrum  we calculate, numerically or
analytically, the derivatives\begin{equation}
\left(\frac{\partial\ln P_{\textrm{obs}}}{\partial 
p_{i}}\right)_{F}\;,\label{eq:deriv}\end{equation}
evaluated at the parameter values of the fiducial model and we obtain for the $i$-th redshift bin all the elements of the Fisher matrix
through \cite{tegmark97}\begin{equation}
F_{ij}=\frac{1}{8\pi^{2}}\int_{-1}^{+1}d\mu\int_{k_{\textrm{min}}}^{k_{\textrm{max}}}dk\
k^{2}\left(\frac{\partial\ln P_{\textrm{obs}}}{\partial 
p_{i}}\frac{\partial\ln P_{\textrm{obs}}}{\partial 
p_{j}}\right)_{F}V_{\textrm{eff}}(k,\mu)\exp[-k^{2}\Sigma_{\perp}^{\,2}-k^{2}\mu^{2}(\Sigma_{\parallel}^{\,2}-\Sigma_{\perp}^{\,2})]\,,\label{eq:fm_elements}\end{equation}
  where \begin{equation}
V_{\textrm{eff}}(k,\mu)=\left[\frac{nP(k,\mu)}{nP(k,\mu)+1}\right]^{2}V_{\textrm{survey}}\,,\label{eq:v_eff}
\end{equation}
is the effective volume of the survey sampled at the scale $k$ along the direction $\mu$.
$V_{\textrm{survey}}$ and $n$ represent
the volume of the survey and the mean number density of galaxies
in each redshift bin.

As a fiducial model  we assume a ``pseudo'' $\Lambda$CDM
with $w_{0}=-0.95$; the differences with the standard $w_{0}=-1.0$  $\Lambda$CDM model are rather small. For example, in the case of  the {\it 
$\gamma$-parameterization}, our fiducial model has $\gamma_{0}=0.545,\,\gamma_{1}=0$ whereas the standard $\Lambda$CDM model has $\gamma_{0}=0.556$, $\gamma_{1}-0.018$~\cite{fu09}.
To summarize, our fiducial model is the same model recommended by the Dark Energy Task Force~\cite{DETF}, i.e.: $\Omega_{m,0}^F=0.25,$ $\Omega_{b,0}^F=0.0445$, 
$\Omega_{k}^F=0$, $h^F=0.7$, $n_{s}^F=1$, $w_{0}^F=-0.95$, $w_{1}^F=0$, $\gamma^F=0.545$, $P_{s}^F=0$.
In addition, we assume that  $\gamma_{1}^F=0$, $\eta^F=0$.
The fiducial values for the redshift dependent parameters are computed in every bin through the standard Friedmann-Robertson-Walker relations

\begin{align}
\frac{H^F(z)}{100 h^F} & \;=\;\left[\Omega_{m,0}^F(1+z)^3+(1-\Omega_{m,0}^F)(1+z)^{3(1+w_0^F)}\right]^{1/2}~,\label{eq:HzF}\\
D^F(z) & \;=\;(1+z)^{-1} \int_0^z\frac{dz'}{H^F(z')}~,\label{eq:DzF}\\
s^F(z) & \;=\;\Omega_m^F(z){\gamma^F}~,\label{eq:szF}\\
G^F(z) & \;=\;\exp\left\{\int_{0}^{z}s^F(z)\frac{dz}{1+z}\right\}~,\label{eq:GzF}\\
\beta^F(z) & \;=\;\frac{\Omega_m^F(z)^{\gamma^F}}{b^F(z)}~,\label{eq:betazF}\\
b^F(z) & \;=\;1\quad \textrm{or}\quad b^F(z) \;=\;\sqrt{1+z}~,\label{eq:bzF}\\
P_s^F(z) & \;=\;0~.\label{eq:P_szF}
\end{align}

At this point our analysis is performed in two ways, according to the choice of $z$-dependent parameters that characterize the power spectrum:

\begin{itemize}
\item{\it{Internal bias method}}.

We assume some fiducial form for $b(z)$ ($z$-dependent or constant) and express the growth function $G(z)$ and the redshift distortion parameter $\beta(z)$ in terms of the growth rate $s$ (see eqs.~(\ref{eq:gengz}), (\ref{eq:beta_sb})). When we compute the derivatives of the spectrum (eq.~(\ref{eq:deriv})), $b(z)$ and $s(z)$ are considered as independent parameters in each redshift bin. In this way we can compute the errors
on $b$ (and $s$) self consistently by marginalizing over
all other parameters.

\item{\it{External bias method}}.

In this case we also assume the same forms for $b(z)$ as in the
{\it Internal bias} case but we do not explicit $G(z)$ and $\beta(z)$ in terms of $s$. The independent parameters are now the product $G(z)\cdot b(z)$ (if we considered them separately, the Fisher matrix would result singular) and $\beta(z)$. In this case we compute the errors over $\beta(z)$ marginalizing over all other parameters. Since we also marginalize over $(G\cdot b)^2$, in this case we cannot estimate the error over $b$ from the Fisher matrix. Thus, in order to obtain the error over $s$ (related to $\beta$ through $s=\beta\cdot b$) with standard error propagation, we need to assume an {}``external'' error for $b(z)$. We allow the relative error $\Delta b/b$ to be either 1\% or 10\%, two values that bracket the ranges of expected errors
contributed by model uncertainties and deviations from linear biasing.

\end{itemize}

\begin{table}
\begin{tabular}{|>{\raggedright}m{6cm}>{\centering}m{2cm}>{\centering}m{4cm}|}
\hline 
Cosmological parameters in $P_{\textrm obs}(z;k,\mu)$  & & Fiducial values\tabularnewline
Reduced total matter density  & $\Omega_{m,0} h^2$ & $0.25\cdot(0.7)^2$\tabularnewline
Reduced baryon density  & $\Omega_{b,0} h^2$ & $0.0445\cdot(0.7)^2$\tabularnewline
Curvature density  & $\Omega_{k}$ & 0\tabularnewline
Hubble constant at present  & $h$ & 0.7\tabularnewline
Primordial fluctuation slope  & $n_{s}$ & 1\tabularnewline
Constant growth index & $\gamma$ & 0.545\tabularnewline
{\it $\gamma$-parameterization} parameters & $\gamma_0$, $\gamma_1$ & 0.545, 0\tabularnewline
{\it $\eta$-parameterization} parameters & $\gamma$, $\eta$ & 0.545, 0\tabularnewline
 & & \tabularnewline
Redshift dependent parameters  & &\tabularnewline
Hubble parameter  & $\log H$ & eq.~(\ref{eq:HzF})\tabularnewline
Angular diameter distance  & $\log D$ & eq.~(\ref{eq:DzF})\tabularnewline
Growth rate  & $\log s$& eq.~(\ref{eq:szF})\tabularnewline
Growth factor  & $\log G$ & eq.~(\ref{eq:GzF})\tabularnewline
Redshift distortion parameter  & $\log\beta$& eq.~(\ref{eq:betazF})\tabularnewline
Shot noise  & $P_{s}$ & 0\tabularnewline
Bias  & $\log b$ & 1, $\sqrt{1+z}$ \tabularnewline
\hline
\end{tabular}

\caption{\label{tab:parameters}
Parameters. }
\end{table}

\section{Modeling the Redshift Survey}\label{sec:survey}

The main goals of next generation redshift surveys will be
to constrain the Dark Energy parameters and  to explore models
alternative to standard Einstein Gravity. For these purposes they will
need to consider very large volumes that encompass
$z\sim 1$, i.e. the epoch at which dark energy  started dominating
the energy budget, spanning a range of epochs large enough to
provide a sufficient leverage to discriminate among
competing models at different redshifts.
The additional requirement is to observe some homogeneous class of objects that are common enough to allow a dense sampling of the underlying mass density field.

As anticipated in the introduction, in this paper we consider as a reference case
 the spectroscopic survey proposed by the EUCLID collaboration~\cite{euclid}.
We stress that our aim is not to focus on this particular redshift 
survey and assess how the constraints on the relevant parameters depends on the survey characteristics in order to optimize future observational strategies.
On the contrary, under the hypothesis that next-generation space-based
all-sky redshift surveys will be similar to the EUCLID spectroscopic survey, we consider the latter as a reference case and estimate how the expected errors on the bias, growth rate, coupling constant and other relevant quantities  will change  when one consider slightly different observational setups.
For this purpose we take advantage of the huge effort made by the EUCLID team to simulate the characteristic of the target objects and compute the expected selection function and detection efficiency of the survey and adopt the same survey parameters presented in~\cite{geach10}.

Here we consider a survey covering a large fraction of the extragalactic
sky ($ |b| \ge 20^{\circ}$), corresponding to $\sim 20000$ deg$^2$
capable to measure a large number of galaxy redshifts out to $z\sim 2$.
A promising observational strategy is to target H$_\alpha$ emitters at near-infrared  wavelengths (which implies $z>0.5$) since they guarantee both relatively dense sampling (the space density of
this population is expected to increase out to $z\sim 2$) and an 
efficient method to measure the redshift of the object.
The limiting flux of the survey should be the tradeoff between
the requirement of minimizing the shot noise, the contamination by
other lines (chiefly among them the [O$_{\rm II}$] line), and that of 
maximizing the so-called efficiency $\varepsilon$, i.e. the fraction of successfully measured redshifts.
To minimize shot noise one should obviously strive for a low flux. Indeed, the authors in~\cite{geach10} found that  a limiting flux  $f_{{\rm H} \alpha} \ge 1\times10^{-16}$  $ \textrm{erg } \textrm{cm}^{-2}  \textrm{s}^{-1}$ would be able to balance shot noise and cosmic variance
out to $z=1.5$. However, simulated observations of mock  H$_\alpha$ galaxy spectra have shown that $\varepsilon$ ranges between 30 \% and 60\% (depending on the redshift) for a limiting flux $f_{{\rm H}_\alpha}\ge 3\times10^{-16}$  $\textrm{erg } \textrm{cm}^{-2}  \textrm{s}^{-1}$~\cite{euclid}. Moreover, contamination from  [O$_{\rm II}$] line drops from 12\% to 1\% when the limiting flux increases from $1\times10^{-16}$ to
$5\times10^{-16}$~\cite{geach10}. Taking all this into account we adopt a conservative choice and consider three different surveys characterized by a limiting flux of 3, 4 and $5\times10^{-16} \, \textrm{erg} \textrm{cm}^{-2}  \textrm{s}^{-1}$.

We use the number density of H$_\alpha$ galaxies at a given redshift, $n(z)$, estimated in~\cite{geach10} using the latest empirical data and obtained by integrating the  H$_\alpha$ luminosity function
above the minimum luminosity set by the limiting flux $L_{{\rm H_
\alpha}, {\rm min.}}= 4 \pi D_L(z)^2 f_{{\rm H}_\alpha}\, $ where $D_L(z)$ is the luminosity 
distance.
To obtain the effective number density one has to account for the 
success rate in measuring galaxy redshifts from  $H_\alpha$ lines. The effective number density is then obtained by multiplying $n(z)$ by the already mentioned efficiency, $\varepsilon$.
In the range of redshifts and fluxes considered in this work the value 
of $\varepsilon$ varies in the interval [30\%, 50\%] (see Fig. A.1.4 of~\cite{euclid}).

In an attempt to bracket current uncertainties in modeling galaxy surveys, we consider the following choices for the survey parameters:

\begin{itemize}

\item {\it Reference case (ref.)}. Limiting flux: $f_{{\rm H} \alpha}\ge 4\times10^{-16}$  
$\textrm{erg } \textrm{cm}^{-2}  \textrm{s}^{-1}$, which gives the galaxy number densities listed in Col.~2 of Tab.~\ref{tab:n_z}. The efficiency is set to $\varepsilon=0.5$. 
\item {\it Optimistic case (opt.)}. Limiting flux: $f_{{\rm H} \alpha}\ge 3\times10^{-16}$  
$\textrm{erg } \textrm{cm}^{-2}  \textrm{s}^{-1}$, which gives the galaxy number densities listed in Col.~1 of Tab.~\ref{tab:n_z}. The efficiency is set to $\varepsilon=0.5$. 
\item {\it Pessimistic case (pess.)}. Limiting flux: $f_{{\rm H} \alpha}\ge 5\times10^{-16}$  
$\textrm{erg } \textrm{cm}^{-2}  \textrm{s}^{-1}$, which gives the galaxy number densities listed in Col.~3 of Tab.~\ref{tab:n_z}. The efficiency is set to $\varepsilon=0.3$. 

\end{itemize}
The total number of observed galaxies ranges from $3\cdot10^7$ (pess.) to $9\cdot10^7$ (opt.).
For all cases we assume that the relative error on the measured 
redshift is $\sigma_z=0.001$, independent of the  limiting flux of the survey.

\begin{table}
\begin{tabular}{|>{\centering}p{2.5cm}|>{\centering}p{2.5cm}|>{\centering}p{2.5cm}|>{\centering}p{2.5cm}|}
\hline 
$z$  & $n_{1}(z)$$\times10^{-3}$  & $n_{2}(z)$$\times10^{-3}$  & $n_{3}(z)$$\times10^{-3}$
\tabularnewline
\hline 
0.5-0.7  & $4.69$  & $3.56$  & $2.8$\tabularnewline
0.7-0.9  & $3.33$  & $2.42$  & $1.84$\tabularnewline
0.9-1.1  & $2.57$  & $1.81$  & $1.33$\tabularnewline
1.1-1.3  & $2.1$   & $1.44$  & $1.03$ \tabularnewline
1.3-1.5  & $1.52$  & $0.99$  & $0.68$ \tabularnewline
1.5-1.7  & $0.92$  & $0.55$  & $0.35$\tabularnewline
1.7-1.9  & $0.54$  & $0.29$  & $0.17$\tabularnewline
1.9-2.1  & $0.31$  & $0.15$  & $0.08$\tabularnewline
\hline
\end{tabular}

\caption{\label{tab:n_z}Expected galaxy number densities in units of ($h/$Mpc)$^{3}$
for EUCLID survey.}

\end{table}

%


\section{Results }\label{sec:results}

In this section we present the main result of the Fisher matrix analysis 
that we split into two sections to stress the different emphasis given in the two approaches.
We note that in all tables below we always quote errors at 68\% probability level and draw in the plots the probability regions at 68\% and/or 95\% (denoted for shortness as 1 and 2$\sigma$ values). Moreover, in all figures, all the parameters that are not shown have been marginalized over or fixed to a fiducial value when so indicated.

\subsection{$s$-parameterization}

This analysis has two main goals: that of figuring out our ability to estimate the
biasing parameter and that of estimating the growth rate with no assumptions on its redshift dependence.
The total number of parameters that enter in the Fisher matrix analysis is 45:
5 parameters that describe the background cosmology
($\Omega_{m,0}h^{2},\Omega_{b,0}h^{2},$ $h$, $n$, $\Omega_{k}$)
plus 5 $z$-dependent parameters specified in 8 redshift bins evenly spaced in the range $z=[0.5,2.1]$. They are $P_{\textrm{s}}(z)$, $D(z)$, $H(z)$, $s(z)$, $b(z)$ in the {\it internal bias} case, while we have $\beta(z)$ and $G(z)\cdot b(z)$ in the place of $s(z)$ and $b(z)$ when we use the {\it external bias} method.

The subsequent analysis depends on the bias method adopted.

\begin{figure}[t]

\begin{centering}
\includegraphics[width=10cm]{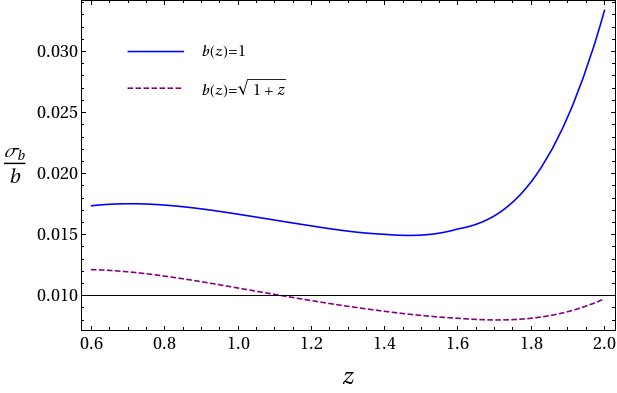} 
\par\end{centering}

\caption{\label{fig:b_err_comp}
Relative errors over the bias parameter as a function of redshift, computed through the Fisher matrix analysis with the ``internal bias'' method for the {\it pessimistic} case. The blue solid line refers to the fiducial bias $b=1$, while the purple dashed line refers to $b=\sqrt{1+z}$. The precision in measuring the bias has a little dependence on the $b(z)$ form: errors are very similar (the discrepancy is less than 1\%) but in the outermost redshift shells (where however is less than 3\%).}
\end{figure}

\begin{figure}[t]

\begin{centering}
\includegraphics[width=12cm]{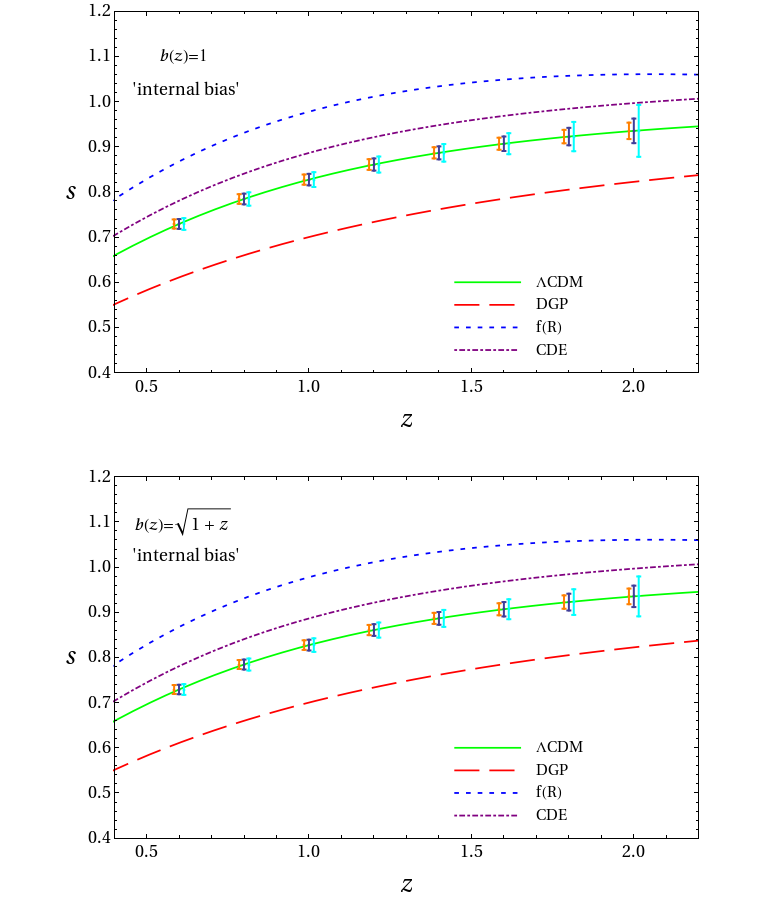} 
\par\end{centering}

\caption{\label{fig:s_err_bint}
Expected constraints on the growth rates
in each redshift bin (using the {}``internal bias'' method). The
upper panel refers to $b=1$, while the lower panel to $b=\sqrt{1+z}$.
For each $z$ the central error bars refer to the {\it Reference case} while
those referring to the {\it Optimistic} and {\it Pessimistic} case have been shifted
by -0.015 and +0.015 respectively. The growth rates for four different
models are also plotted: $\Lambda$CDM (green solid curve), flat DGP
(red dashed curve), $f(R)$ model (blue dotted curve) and a model
with coupling between dark energy and dark matter (purple, dot-dashed
curve). In this case it will be possible to distinguish these models
with next generation data.}
\end{figure}

\begin{figure}[t]

\begin{centering}

\includegraphics[width=12cm]{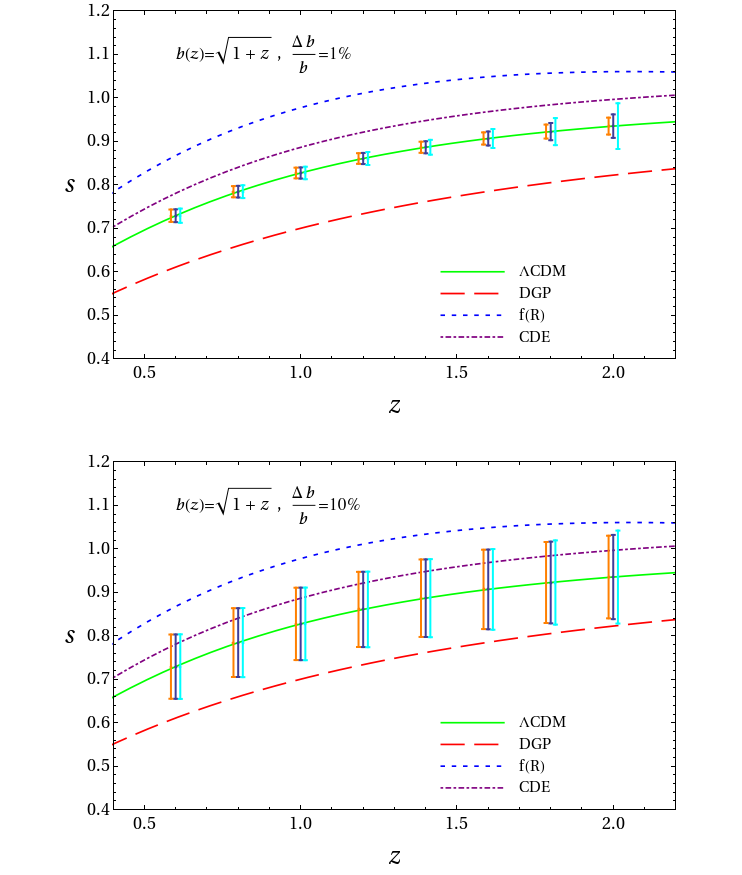}
\end{centering}
\caption{\label{fig:s_err}
Expected constraints on the growth rates in each redshift bin (using
the {}``external bias'' method), assuming for the bias a relative
error of 1\% (upper panel) and 10\% (lower panel). For each
$z$ the central error bars refer to the {\it Reference case} while those
referring to the {\it Optimistic} and {\it Pessimistic} case have been shifted
by -0.015 and +0.015 respectively. The growth rates for four different
models are also plotted: $\Lambda$CDM (green solid curve), flat DGP
(red dashed curve), $f(R)$ model (blue dotted curve) and a model
with coupling between dark energy and dark matter (purple, dot-dashed
curve). Even in the case of large errors (10\%) for the bias it will
be possible to distinguish among three of these models with next generation data.} 
\end{figure}

\begin{itemize}

\item In case of the {\it internal bias} method, the fiducial growth function $G(z)$ in the $(i+1)$-th redshift bin is evaluated from a step-wise, constant growth rate $s(z)$ as
\begin{equation}
G(z)=\exp\left\{\int_{0}^{z}s(z)\frac{dz}{1+z}\right\}=\prod_{i}\left(\frac{1+z_{i}}{1+z_{i-1}}\right)^{s_{i}}\left(\frac{1+z}{1+z_{i}}\right)^{s_{i+1}}\,.\label{eq:gengz}\end{equation}
To obtain the errors on $s_{i}$ and $b_{i}$ we compute the elements of the Fisher matrix and marginalize over all other parameters. In this case one is able to obtain, self-consistently, the error on the bias and on the growth factor at different redshifts, as detailed in Tab.~\ref{tab:sigma_bias_bint} and Tab.~\ref{tab:sigma_s_bint} respectively.

Tab.~\ref{tab:sigma_bias_bint} illustrates one important result: through the analysis of
the redshift-space galaxy power spectrum in a next-generation
EUCLID-like survey, it will be possible to measure galaxy biasing
in $\Delta z=0.2$ redshift bins with less than 3.5\% error, provided
that the bias function is independent on scale. 
This fact can be appreciated in Fig.~\ref{fig:b_err_comp} in which we show the expected relative error as a function of redshift for both $b(z)$ functions and for the survey {\it Pessimistic case}. Errors are very similar in all but 
the outermost redshift shells. We show the  {\it Pessimistic case}
since  with a more favorable survey configuration, like the
{\it Reference case}, the errors would be almost identical.
In addition we find that the precision in
measuring the bias has a little dependence on the $b(z)$ form.
 The largest discrepancy between the
  $b(z)=1$ and  $b(z)=\sqrt{1+z}$ cases is $\sim 3 \%$
  and refers to the expected errors on the growth rate
  at $z=2$ in the   {\it Pessimistic case}. Differences are typically much smaller
  for all other parameters or, for $s(z)$ at lower redshifts or with
  a more favorable survey setup.
Given the robustness of the results on the choice of $b(z)$  
in the following we  only consider the $b(z)=\sqrt{1+z}$ case.

In Fig.~\ref{fig:s_err_bint} we show the errors on
the growth rate $s$ as a function of redshift,
overplotted to our fiducial $\Lambda$CDM (green solid
curve).  The three sets of error bars are plotted in correspondence of
the 8 redshift bins and refer (from left to right) to the {\it Optimistic, 
Reference} and {\it Pessimistic}
cases, respectively. The other curves show the expected growth rate in
three alternative cosmological models:  flat DGP (red dashed curve),   
$f(R)$ (blue dotted curve) and CDE (purple, dot-dashed curve).
This plot clearly illustrates the ability of next generation surveys to 
distinguish between alternative models, even in the less favorable choice of survey parameters.

\item In case of the {\it external bias} method we marginalize over the 
overall amplitude $(G\cdot b)^{2}$.
Since, in this case, we cannot find errors self-consistently, we assume
that bias has been determined {\it a priori} with errors per redshift bin
of 1\% and 10\%, two values that should bracket the expected range of 
uncertainties.
We note that the {\it external bias} method can be considered more conservative, especially in the case of large errors although we see no obvious reason why it should be preferred to the  
{\it internal bias} method that seems to provide similar results.
Indeed, the errors on $s$ relative to the 1\% bias error listed
in  Table~\ref{tab:sigma_beta_s_b1} are quite similar to those of
the {\it internal bias} case. As expected, errors on $s$ increase 
significantly when the bias is known with 10\% accuracy rather than 1\%. However, even in this case, one keeps the ability of distinguishing between most of the competing cosmological models at 1$\sigma$ level, as shown
in Fig.~\ref{fig:s_err}.

\end{itemize}

\begin{table}
\begin{tabular}{|>{\centering}p{1.3cm}|>{\centering}p{1.3cm}|>{\centering}p{1.3cm}|>{\centering}p{1cm}|>{\centering}p{1.3cm}|>{\centering}p{1.3cm}|>{\centering}p{1.3cm}|}
\hline 
\multicolumn{3}{|c|}{$b(z)=1$ (internal)} &  & \multicolumn{3}{c|}{$b(z)=\sqrt{1+z}$ (internal)}\tabularnewline
\hline
\multicolumn{1}{|>{\centering}p{1.3cm}}{} & \multicolumn{1}{>{\centering}p{1.3cm}}{$\sigma_{b}$} &  & $z$  & \multicolumn{1}{>{\centering}p{1.3cm}}{} & \multicolumn{1}{>{\centering}p{1.3cm}}{$\sigma_{b}$} & \tabularnewline
\cline{2-2} \cline{3-3} \cline{4-4} \cline{5-5} \cline{7-7} 
\hline 
ref.  & opt.  & pess.  &  & ref.  & opt.  & pess.\tabularnewline
\hline
0.014  & 0.012  & 0.017  & 0.6  & 0.012  & 0.011  & 0.015\tabularnewline
0.014  & 0.012  & 0.017  & 0.8  & 0.012  & 0.011  & 0.016\tabularnewline
0.013  & 0.012  & 0.017  & 1.0  & 0.012  & 0.011  & 0.015\tabularnewline
0.012  & 0.011  & 0.016  & 1.2  & 0.011  & 0.010  & 0.014\tabularnewline
0.012  & 0.011  & 0.015  & 1.4  & 0.011  & 0.010  & 0.013\tabularnewline
0.012  & 0.010  & 0.025  & 1.6  & 0.010  & 0.010  & 0.013\tabularnewline
0.012  & 0.010  & 0.019  & 1.8  & 0.010  & 0.010  & 0.014\tabularnewline
0.016  & 0.011  & 0.033  & 2.0  & 0.011  & 0.010  & 0.017\tabularnewline
\hline
\end{tabular}

\caption{\label{tab:sigma_bias_bint}
$1\sigma$ marginalized errors for the bias in each redshift
bin obtained with the {}``internal bias'' method. }
\end{table}

The main results of this section can be summarized as follows.
\begin{enumerate}

\item The ability of measuring the biasing function is not too sensitive
to the characteristic of the survey ($b(z)$ can be constrained to within 
1.5\% in the {\it Optimistic} scenario and up to 3.5\% in the {\it Pessimistic} one) provided
that the bias function is independent on scale. Moreover, the precision in
measuring the bias has a very little dependence on the $b(z)$ form.

\item The growth rate $s$ can be estimated to within 1-3\%
in each bin for the {\it Reference case} survey with
no need of estimating the bias function $b(z)$ from some
dedicated, independent analysis using higher order statistics~\cite{verde02}
or full-PDF analysis~\cite{sigad00}.

\item If the the bias were measured to within 1\% in each slice, then
the  error over $s$ would be very similar (just 1-2\% larger) to that
obtained by the internal estimate of $b(z)$.

\item The estimated errors on $s$ depend weakly on the fiducial
model of $b(z)$.
\end{enumerate}

\begin{table}
\begin{tabular}{|>{\centering}p{1.3cm}|>{\centering}p{1.3cm}|>{\centering}p{1.3cm}|>{\centering}p{1cm}|>{\centering}p{1cm}|>{\centering}p{1.3cm}|>{\centering}p{1.3cm}|>{\centering}p{1.3cm}|}
\hline 
\multicolumn{3}{|c|}{$b(z)=1$ (internal)} & & & \multicolumn{3}{c|}{$b(z)=\sqrt{1+z}$ (internal)}\tabularnewline
\hline
\multicolumn{1}{|>{\centering}p{1.3cm}}{} & \multicolumn{1}{>{\centering}p{1.3cm}}{$\sigma_{s}$} &  & $z$ & $s^F$ & \multicolumn{1}{>{\centering}p{1.3cm}}{} & \multicolumn{1}{>{\centering}p{1.3cm}}{$\sigma_{s}$} & \tabularnewline
\hline 
ref.  & opt.  & pess.  & & & ref.  & opt.  & pess.\tabularnewline
\hline
0.011  & 0.010  & 0.013  & 0.6 & 0.73  & 0.010  & 0.010  & 0.012\tabularnewline
0.011  & 0.010  & 0.015  & 0.8 & 0.78  & 0.011  & 0.010  & 0.013\tabularnewline
0.013  & 0.011  & 0.016  & 1.0 & 0.83  & 0.012  & 0.010  & 0.015\tabularnewline
0.013  & 0.012  & 0.018  & 1.2 & 0.86  & 0.013  & 0.011  & 0.017\tabularnewline
0.014  & 0.012  & 0.019  & 1.4 & 0.89  & 0.014  & 0.012  & 0.019\tabularnewline
0.016  & 0.013  & 0.023  & 1.6 & 0.91  & 0.016  & 0.013  & 0.022\tabularnewline
0.019  & 0.015  & 0.032  & 1.8 & 0.92  & 0.018  & 0.015  & 0.028\tabularnewline
0.027  & 0.018  & 0.057  & 2.0 & 0.93  & 0.024  & 0.017  & 0.044\tabularnewline
\hline
\end{tabular}

\caption{\label{tab:sigma_s_bint}
$1\sigma$ marginalized errors for the growth rates in each redshift
bin (Fig.~\ref{fig:s_err_bint}) obtained with the {}``internal bias'' method.} 
\end{table}

%
\begin{table}
\begin{tabular}{|>{\centering}p{1cm}|>{\centering}p{1cm}|>{\centering}p{1cm}|>{\centering}p{1cm}|>{\centering}p{1cm}|>{\centering}p{1cm}|>{\centering}p{1cm}|>{\centering}p{1cm}|>{\centering}p{1cm}|>{\centering}p{1cm}|>{\centering}p{1cm}|>{\centering}p{1cm}|>{\centering}p{1cm}|}
\hline 
\multicolumn{6}{|c|}{$b(z)=1$ (external)} &  & \multicolumn{6}{c|}{$b(z)=\sqrt{1+z}$ (external)}\tabularnewline
\hline
\hline 
\multicolumn{3}{|c|}{$\Delta b/b=1\%$} & \multicolumn{3}{c|}{$\Delta b/b=10\%$} & $z$  & \multicolumn{3}{c|}{$\Delta b/b=1\%$} & \multicolumn{3}{c|}{$\Delta b/b=10\%$}\tabularnewline
\hline 
\multicolumn{1}{|>{\centering}p{1cm}}{} & \multicolumn{1}{>{\centering}p{1cm}}{$\sigma_{s}$} &  & \multicolumn{1}{>{\centering}p{1cm}}{} & \multicolumn{1}{>{\centering}p{1cm}}{$\sigma_{s}$} &  &  & \multicolumn{1}{>{\centering}p{1cm}}{} & \multicolumn{1}{>{\centering}p{1cm}}{$\sigma_{s}$} &  & \multicolumn{1}{>{\centering}p{1cm}}{} & \multicolumn{1}{>{\centering}p{1cm}}{$\sigma_{s}$} & \tabularnewline
\hline 
ref.  & opt.  & pess.  & ref.  & opt.  & pess.  &  & ref.  & opt.  & pess.  & ref.  & opt.  & pess.\tabularnewline
\hline 
0.014  & 0.013  & 0.015  & 0.074  & 0.074  & 0.074  & 0.6  & 0.015  & 0.014  & 0.016  & 0.074  & 0.074  & 0.074\tabularnewline
0.012  & 0.012  & 0.014  & 0.079  & 0.079  & 0.079  & 0.8  & 0.013  & 0.013  & 0.015  & 0.079  & 0.079  & 0.079\tabularnewline
0.012  & 0.011  & 0.014  & 0.083  & 0.083  & 0.083  & 1.0  & 0.013  & 0.012  & 0.014  & 0.083  & 0.083  & 0.083\tabularnewline
0.012  & 0.012  & 0.015  & 0.086  & 0.086  & 0.087  & 1.2  & 0.013  & 0.012  & 0.015  & 0.086  & 0.087  & 0.087\tabularnewline
0.014  & 0.012  & 0.018  & 0.089  & 0.089  & 0.089  & 1.4  & 0.014  & 0.013  & 0.027  & 0.089  & 0.089  & 0.089\tabularnewline
0.017  & 0.014  & 0.025  & 0.092  & 0.091  & 0.094  & 1.6  & 0.016  & 0.014  & 0.022  & 0.092  & 0.091  & 0.093\tabularnewline
0.023  & 0.017  & 0.042  & 0.095  & 0.093  & 0.101  & 1.8  & 0.019  & 0.016  & 0.031  & 0.094  & 0.093  & 0.097\tabularnewline
0.036  & 0.023  & 0.082  & 0.099  & 0.096  & 0.124  & 2.0  & 0.027  & 0.019  & 0.052  & 0.097  & 0.095  & 0.107\tabularnewline
\hline
\end{tabular}

\caption{\label{tab:sigma_beta_s_b1}
$1\sigma$ marginalized errors for the growth rates in each redshift
bin (Fig.~\ref{fig:s_err}) obtained with the {}``external bias'' method. }
\end{table}

\subsection{Other parameterizations.}
In this section we assess the ability of estimating $s(z)$ when it is
expressed in one of the  parametrized forms described in Section~\ref{subsec:parameterizations}.
More specifically, we focus on the ability of determining $\gamma_0$
and $\gamma_1$, in the context of the {\it $\gamma$-parameterization}
and $\gamma$, $\eta$ in the {\it $\eta$-parameterization}.
In both cases the Fisher matrix elements have been estimated
by expressing the growth factor as
\begin{eqnarray}
G(z) & = & 
\delta_{0}\exp\left[(1+\eta)\int_{0}^{z}\Omega_{m}(z^{\prime})^{\gamma(z)}\frac{dz^{\prime}}{1+z^{\prime}}\right]
\, ,
\label{eq:growth_fact_def_gam}\end{eqnarray}
where  for the {\it $\gamma$-parameterization} we fix $\eta=0$.
In this section we adopt the {\it internal bias} approach and assume
that $b(z)=\sqrt{1+z}$ since, as we have checked, in the case
of $b(z)=1$ one obtains  very similar results.

\begin{figure}[t]

\begin{centering}
\includegraphics[width=8cm]{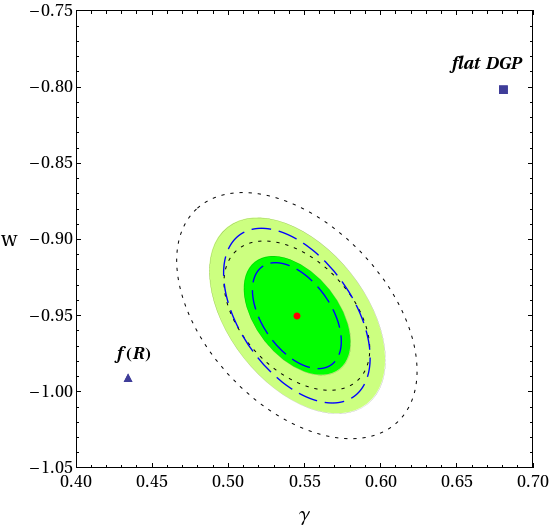} 
\par\end{centering}

\caption{\label{fig:gam_w_b1_dgp}
$\gamma$-parameterization. 1 and 2$\sigma$ marginalized probability
regions for constant $\gamma$ and $w$: the green (shaded) regions
are relative to the {\it Reference case}, the blue long-dashed ellipses
to the {\it Optimistic case}, while the black short-dashed ellipses are
the probability regions for the {\it Pessimistic case}. The red dot marks the fiducial model; two alternative models are also indicated for comparison.} 
\end{figure}

\begin{figure}[t]

\begin{centering}
\includegraphics[width=16cm]{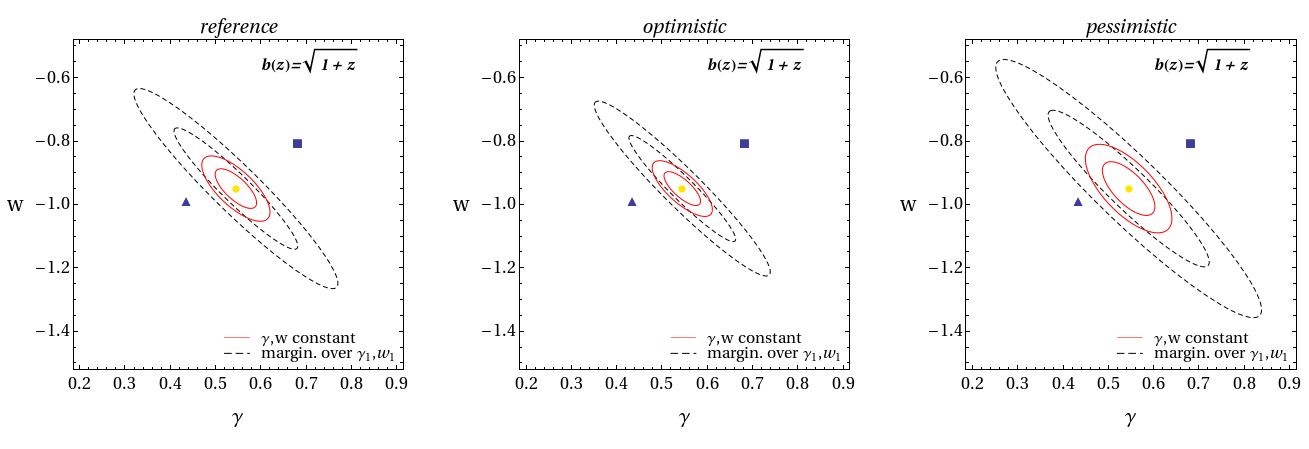} 
\par\end{centering}

\caption{\label{fig:gam_w_margover_gam1w1}
$\gamma$-parameterization. 1 and 2$\sigma$ marginalized probability
regions obtained assuming constant $\gamma$ and $w$ (red solid curves)
or assuming the parameterizations (\ref{eq:gam_CPL}) and (\ref{eq:w_CPL})
and marginalizing over $\gamma_{1}$ and $w_{1}$ (black dashed curves);
marginalized error values are in columns $\sigma_{{\gamma}_{\textrm{marg},1}}$,
$\sigma_{{w}_{\textrm{marg},1}}$ of Tab.~\ref{tab:sigma_gam_w_marg}. Yellow dots
represent the fiducial model, the triangles a $f(R)$ model and the squares mark the flat DGP.}
\end{figure}

\begin{figure}[t]

\begin{centering}
\includegraphics[width=8cm]{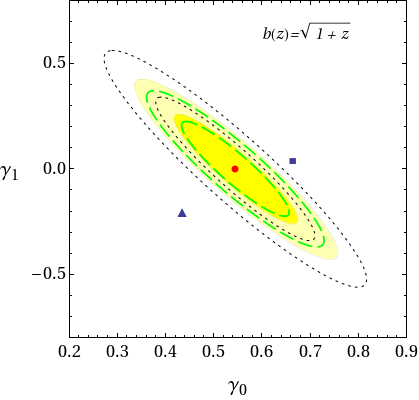} 
\par\end{centering}

\caption{\label{fig:gam0gam1}
$\gamma$-parameterization. 1 and 2$\sigma$ marginalized probability
regions for the parameters $\gamma_{0}$ and $\gamma_{1}$, relative
to the {\it Reference case} (shaded yellow regions), to the {\it Optimistic case}
(green long-dashed ellipses), and to the {\it Pessimistic case} (black dotted
ellipses). Red dots represent the fiducial model, blue squares
mark the DGP while triangles stand for the $f(R)$ model. Then, in the case of $\gamma$-parameterization, one
could distinguish these three models (at 95\% probability). }
\end{figure}

\begin{figure}[t]

\begin{centering}
\includegraphics[width=10cm]{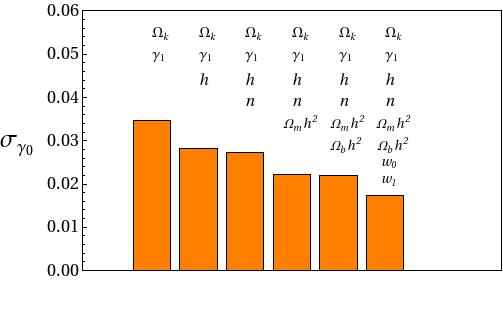} 
\par\end{centering}

\caption{\label{fig:histogram}
The bars represent the errors on the growth index $\gamma_{0}$ obtained using the {\it $\gamma$-parameterization} and fixing an increasing number of cosmological parameters as indicated over each bar and marginalizing over the others. The progressive increase in the number of fixed parameters reflects in a decrease of the error.}
\end{figure}

\begin{figure}[t]

\begin{centering}
\includegraphics[width=16cm]{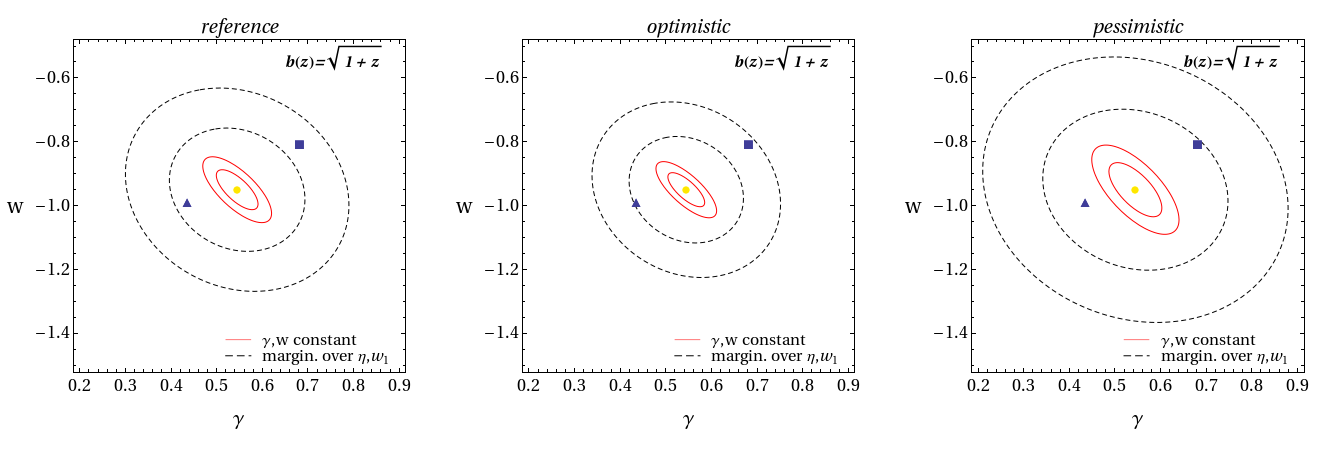} 
\par\end{centering}

\caption{\label{fig:gam_w_margover_etaw1}
$\eta$-parameterization. 1 and 2$\sigma$ marginalized probability
regions obtained assuming constant $\gamma$ and $w$ (red solid curves)
or assuming the parameterizations (\ref{eq:eta_paramet}) and (\ref{eq:w_CPL})
and marginalizing over $\eta$ and $w_{1}$ (black dashed curves);
marginalized error values are in columns $\sigma_{{\gamma}_{\textrm{marg},2}}$,
$\sigma_{{w}_{\textrm{marg},2}}$ of Tab.~\ref{tab:sigma_gam0_gam1}. Yellow dots
represent the fiducial model, the triangles stand for a $f(R)$ model and the squares mark the flat DGP. }
\end{figure}

\begin{figure}[t]
\!\includegraphics[height=7.6cm]{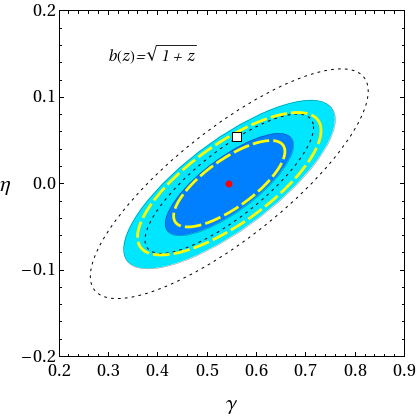}\includegraphics[height=7.8cm]{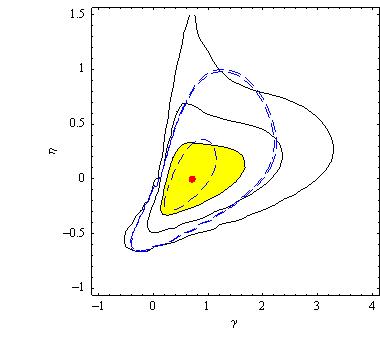}

\caption{\label{fig:gamma_eta_new_past}
$\eta$-parameterization. Left panel: 1 and 2$\sigma$ marginalized probability
regions for the parameters $\gamma$ and $\eta$ in eq.~(\ref{eq:eta_paramet})
relative to the reference case (shaded blue regions), to the optimistic
case (yellow long-dashed ellipses) and to the pessimistic case (black
short-dashed ellipses). The red dot marks the fiducial model while the square represents the coupling model. Right panel: present constraints on $\gamma$ and $\eta$ computed
through a full likelihood method (here the red dot marks the likelihood
peak) \protect \cite{diporto08}; 
long-dashed contours are obtained assuming a prior for $\Omega_{m,0}$. }
\end{figure}

\begin{itemize}

\item {\it $\gamma$-parameterization}.
We start by considering the  case
of constant $\gamma$ and $w$ in which we set $\gamma=\gamma^F=0.545$ and $w=w^F=-0.95$. As we will discuss in the next Section, this simple case will allow us to cross-check our results with those in the literature.
In  Fig.~\ref{fig:gam_w_b1_dgp} we show the marginalized probability 
regions, at 1 and 2$\sigma$ levels, for  $\gamma$ and $w$.
The regions with different shades of green illustrates the
{\it Reference case} for the survey whereas the
blue long-dashed and the black short-dashed ellipses
refer to the {\it Optimistic} and {\it Pessimistic} cases, respectively.
Errors on $\gamma$ and $w$ are listed in Tab.~\ref{tab:sigma_gam_w}
together with the corresponding figures of merit [FOM]
defined to be the squared inverse of the Fisher matrix determinant and therefore equal to the inverse of the
product of the errors in the pivot point, see~\cite{DETF}.
Contours are centered on the fiducial model. The blue triangle and
the blue square represent the  flat DGP and the $f(R)$ models' predictions, respectively.
It is clear that, in the case of constant $\gamma$ and $w$, the
measurement of the growth rate in a EUCLID-like survey will
allow us to discriminate among these models. These results have been 
obtained by fixing the curvature to its fiducial value $\Omega_k=0$. If instead, we consider 
curvature as a free parameter and marginalize over, the errors on $\gamma$ and $w$ increase significantly, as shown in Table~\ref{tab:sigma_gam_w_omk_marg}, and yet the precision is good enough to distinguish the different models.
For completeness, we also computed the fully marginalized errors over the other Cosmological parameters for the reference survey, given in Tab.~\ref{tab:cosm_par_errors}.

As a second step we considered the case in which
$\gamma$ and $w$ evolve with redshift according to
eqs.~(\ref{eq:gam_CPL}) and~(\ref{eq:w_CPL})
and then we marginalize over the parameters $\gamma_{1}$, $w_{1}$ and  
$\Omega_k$.
The marginalized probability contours
are shown  in Fig.~\ref{fig:gam_w_margover_gam1w1} in which we have shown
the three survey setups in three different panels to avoid overcrowding.
Dashed contours refer to the $z$-dependent parameterizations while red, continuous
contours refer to the case of  constant $\gamma$ and $w$ obtained after
marginalizing over $\Omega_k$.
Allowing for time dependency increases the size of the confidence ellipses since the Fisher matrix analysis now accounts for the additional uncertainties in the extra-parameters $\gamma_{1}$ and $w_{1}$; marginalized error values are in columns $\sigma_{{\gamma}_{\textrm{marg},1}}$,
$\sigma_{{w}_{\textrm{marg},1}}$ of Tab.~\ref{tab:sigma_gam_w_marg}.  We note, however, that errors are still small enough to distinguish the fiducial model from the $f(R)$ and DGP scenarios.

We have also projected the marginalized ellipses for the parameters $\gamma_{0}$ and $\gamma_{1}$  and calculated their marginalized
errors and figures of merit, which are reported in 
Tab.~\ref{tab:sigma_gam0_gam1}.
The corresponding uncertainties contours are shown in
Fig.~\ref{fig:gam0gam1}. Once again we overplot the expected values
in the $f(R)$ and DGP scenarios to stress the fact that
one is expected to be able to distinguish among competing models, 
irrespective on the survey's precise characteristics.

As a further test we have estimated how the errors on
$\gamma_0$ depend on the number of parameters
explicitly involved in the Fisher matrix analysis.
Fig.~\ref{fig:histogram} shows the expected 1$\sigma$ errors on
$\gamma$ (Y-axis) as a function of the number of parameters
that are fixed when computing the element of the Fisher matrix
(the different combinations of the parameters are shown on the
top of the histogram elements). We see that error estimates can
decrease up to $\sim 50$ \% when parameters are fixed to
some fiducial value, or are determined independently.

\item {\it $\eta$-parameterization}.

We have repeated the same analysis as for
the {\it $\gamma$-parameterization} taking into account the 
possibility of coupling between DE and DM
i.e. we have
modeled the growth factor according to eq.~(\ref{eq:eta_paramet}) and 
the dark energy equation of state as in eq.~(\ref{eq:w_CPL}) and 
marginalized over all parameters, including $\Omega_k$. The marginalized errors are shown in columns $\sigma_{{\gamma}_{\textrm{marg},2}}$,
$\sigma_{{w}_{\textrm{marg},2}}$ of Tab.~\ref{tab:sigma_gam_w_marg} and the significance contours are shown in the three panels of Fig.~\ref{fig:gam_w_margover_etaw1} which is analogous to Fig.~\ref{fig:gam_w_margover_gam1w1}. The uncertainty ellipses are now larger than in the case of the {\it $\gamma$-parameterization} and show that DGP and $f(R)$ models could be rejected at $>1\sigma$ level only if the redshift survey parameter will be more favorable than in the {\it Pessimistic case}.

Marginalizing over all other parameters we can compute the uncertainties in
the $\gamma$ and $\eta$ parameters, as listed in Tab.~\ref{tab:sigma_gam_eta}.
The relative confidence ellipses are shown in the left panel of Fig.~\ref{fig:gamma_eta_new_past}.
This plot shows that next generation EUCLID-like surveys will be able to 
distinguish the reference model with no coupling (central, red dot) to the CDE model proposed by~\cite{amendola03} (white square) only at the 1-1.5 $\sigma$ level.

\end{itemize}

%



%

\begin{table}
\begin{tabular}{|>{\centering}p{2.5cm}|>{\centering}p{2cm}|>{\centering}p{2cm}|c|>{\centering}p{2cm}|}
\hline 
  & case  & $\sigma_{\gamma}$  & \multicolumn{1}{>{\centering}p{2cm}|}{$\sigma_{w}$} & FOM\tabularnewline
\hline
\hline 
$b=\sqrt{1+z}$ & ref.  & 0.02  & 0.02  & 2115\tabularnewline
  & opt.  & 0.019  & 0.019  & 2806\tabularnewline
$\Omega_{k}$ fixed & pess.  & 0.03  & 0.03  & 1296\tabularnewline
\hline
\end{tabular}

\caption{\label{tab:sigma_gam_w}Numerical values for 1$\sigma$ constraints
on parameters in Fig.~\ref{fig:gam_w_b1_dgp} and figures of merit. Here we have fixed $\Omega_{k}$ to its fiducial value.}

\end{table}




%

\begin{table}
\begin{tabular}{|>{\centering}p{2.5cm}|>{\centering}p{2cm}|>{\centering}p{2cm}|c|>{\centering}p{2cm}|}
\hline 
bias  & case  & $\sigma_{\gamma}$  & \multicolumn{1}{>{\centering}p{2cm}|}{$\sigma_{w}$} & FOM\tabularnewline
\hline
\hline 
 & ref.  & 0.03  & 0.04  & 1179\tabularnewline
$b=\sqrt{1+z}$  & opt.  & 0.03  & 0.03  & 1568\tabularnewline
 & pess.  & 0.04  & 0.05  & 706\tabularnewline
\hline
\end{tabular}

\caption{\label{tab:sigma_gam_w_omk_marg}Numerical values for 1$\sigma$ constraints
on parameters $\gamma$ and $w$ (no parameterization), relative to the red ellipses in Figs~\ref{fig:gam_w_margover_gam1w1},~\ref{fig:gam_w_margover_etaw1}  and figures of merit. Here we have marginalized over $\Omega_{k}$.}

\end{table}

\begin{table}
\begin{tabular}{|>{\centering}p{2.5cm}|>{\centering}p{1cm}|>{\centering}p{1cm}|>{\centering}p{1cm}|>{\centering}p{1cm}|c|>{\centering}p{1cm}|}
\hline 
  & case & $\sigma_{h}$ & $\sigma_{\Omega_m h^2}$  & $\sigma_{\Omega_b h^2}$ & $\sigma_{\Omega_k}$ & $\sigma_{n_s}$\tabularnewline
\hline
$b=\sqrt{1+z}$ & ref.  & 0.024  & 0.008  & 0.002 & 0.01 & 0.02\tabularnewline
\hline
\end{tabular}

\caption{\label{tab:cosm_par_errors}Numerical values for marginalized 1$\sigma$ constraints on Cosmological parameters using constant $\gamma$ and $w$.}

\end{table}
\begin{table}
\begin{tabular}{|>{\centering}p{2.5cm}|>{\centering}p{1.5cm}|>{\centering}p{1.5cm}|c|>{\centering}p{1.5cm}||>{\centering}p{1.5cm}|>{\centering}p{1.5cm}|>{\centering}p{1.5cm}|}
\hline 
bias  & case  & $\sigma_{\gamma_{marg,1}}$  & \multicolumn{1}{>{\centering}p{1.5cm}|}{$\sigma_{w_{marg,1}}$} & FOM  & $\sigma_{\gamma_{marg,2}}$  & $\sigma_{w_{marg,2}}$  & FOM\tabularnewline
\hline
\hline 
 & ref.  & 0.08  & 0.11  & 241.1  & 0.09  & 0.11  & 104.1\tabularnewline
$b=\sqrt{1+z}$  & opt.  & 0.07  & 0.09  & 323.5  & 0.07  & 0.09  & 138.2\tabularnewline
 & pess.  & 0.11  & 0.14  & 142.5  & 0.11  & 0.15  & 61.6\tabularnewline
\hline
\end{tabular}

\caption{\label{tab:sigma_gam_w_marg}1$\sigma$ marginalized errors for parameters
$\gamma$ and $w$ expressed through $\gamma$ and $\eta$ parameterizations.
Columns $\gamma_{0,marg1},w_{0,marg1}$ refer to marginalization over
$\gamma_{1},w_{1}$ (Fig.~\ref{fig:gam_w_margover_gam1w1}) while
columns $\gamma_{0,marg2},w_{0,marg2}$ refer to marginalization over
$\eta,w_{1}$ (Fig.~\ref{fig:gam_w_margover_etaw1}).}

\end{table}

\begin{table}
\begin{tabular}{|>{\centering}p{2.5cm}|>{\centering}p{2cm}|>{\centering}p{2cm}|c|>{\centering}p{2cm}|}
\hline 
bias  & case  & $\sigma_{\gamma_{0}}$  & \multicolumn{1}{>{\centering}p{2cm}|}{$\sigma_{\gamma_{1}}$} & FOM\tabularnewline
\hline
\hline 
 & ref.  & 0.08  & 0.17  & 178.4\tabularnewline
$b=\sqrt{1+z}$  & opt.  & 0.07  & 0.15  & 227.5\tabularnewline
 & pess.  & 0.11  & 0.22  & 112.4\tabularnewline
\hline
\end{tabular}

\caption{\label{tab:sigma_gam0_gam1}Numerical values for 1$\sigma$ constraints
on parameters in Fig.~\ref{fig:gam0gam1} and figures of merit.}

\end{table}

%



%

\begin{table}
\begin{tabular}{|>{\centering}p{2.5cm}|>{\centering}p{2cm}|>{\centering}p{2cm}|c|>{\centering}p{2cm}|}
\hline 
bias  & case  & $\sigma_{\gamma}$  & \multicolumn{1}{>{\centering}p{2cm}|}{$\sigma_{\eta}$} & FOM\tabularnewline
\hline
\hline 
 & ref.  & 0.08  & 0.04  & 464.1\tabularnewline
$b=\sqrt{1+z}$  & opt.  & 0.07  & 0.03  & 608.2\tabularnewline
 & pess.  & 0.11  & 0.05  & 280.3\tabularnewline
\hline
\end{tabular}

\caption{\label{tab:sigma_gam_eta}Numerical values for 1$\sigma$ constraints
on parameters in Fig.~\ref{fig:gamma_eta_new_past} and figures of merit.}

\end{table}

\begin{figure}[t]

\begin{centering}
\includegraphics[width=8cm]{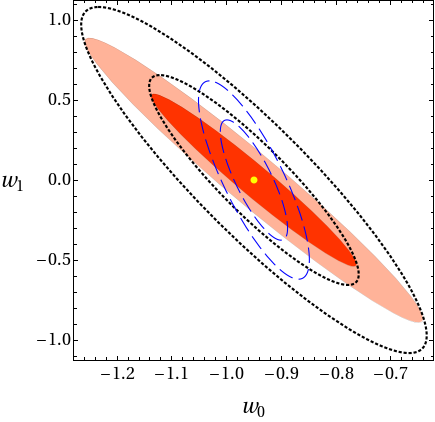} 
\par\end{centering}

\caption{\label{fig:w0_w1}
Errors on the equation of state. 1 and 2$\sigma$ marginalized probability
regions for the parameters $w_{0}$ and $w_{1}$, relative
to the reference case and constant bias $b=1$. The blue dashed ellipses are obtained fixing $\gamma_{0}, \gamma_{1}$ to their fiducial values and marginalizing over all the other parameters; for the red shaded ellipses instead, we also marginalize over $\gamma_{0}, \gamma_{1}$ but we fix $\Omega_{k}=0$. Finally, the black dotted ellipses are obtained marginalizing over all parameters but $w_{0}$ and $w_{1}$. The progressive increase in the number of parameters reflects in a widening of the ellipses with a consequent decrease in the figures of merit (see Tab.~\ref{tab:w0_w1}). }
\end{figure}




Finally, in order to explore the dependence on the number of parameters and to compare our results to previous works, we also draw the confidence ellipses for $w_0$, $w_1$ with three different methods: $i$) fixing $\gamma_{0}, \gamma_{1}$ to their fiducial values and marginalizing over all the other parameters; $ii$) marginalizing over all parameters plus $\gamma_{0}, \gamma_{1}$ but fixing $\Omega_k$; $iii$) marginalizing over all parameters but $w_0$, $w_1$. As one can see in Fig.~\ref{fig:w0_w1} and Tab.~\ref{tab:w0_w1} this progressive increase in the number of marginalized parameters reflects in a widening of the ellipses with a consequent decrease in the figures of merit. These results are in agreement with those of other authors (e.g.~\cite{wang10,majerotto11}).

The results obtained this Section can be summarized as follows.
\begin{enumerate}

\item If both $\gamma$ and $w$ are assumed to be constant
and setting  $\Omega_k=0$
then, a redshift survey described by our {\it Reference case}
will be able to constrain these parameters to within 4\% and 2\%,
respectively.

\item Marginalizing over $\Omega_{k}$ degrades these constraints to 5.5\% and 4\% respectively.

\item If $w$ and $\gamma$ are considered redshift-dependent and
parametrized according to eqs~(\ref{eq:gam_CPL}) and~(\ref{eq:w_CPL})
then the errors on $\gamma_{0}$ and $w_{0}$ obtained after
marginalizing over  $\gamma_{1}$ and  $w_{1}$ increase by a
factor $\sim 4$, 5, i.e. we expect to measure $\gamma_{0}$ and $w_{0}$
with a precision of 13-15\% and 11-14\% respectively, where the interval reflects the uncertainties in the characteristic of the survey. With this precision we will
be able to distinguish the fiducial model from the  DGP and $f(R)$
scenarios with more than 2$\sigma$ significance.

\item The ability to discriminate these models with a significance above
2$\sigma$ is confirmed by the confidence contours drawn in the
$\gamma_{0}$-$\gamma_{1}$ plane, obtained after marginalizing over all other parameters.

\item If we allow for a coupling between dark matter and dark  energy,
and  we marginalize over $\eta$ rather than over $\gamma_{1}$,
then the errors on $\gamma_{0}$ and $w_{0}$ are almost identical
to those obtained in the case of the {\it $\gamma$-parameterization}.
However, our ability in separating the fiducial model from the
CDE model is significantly hampered: the confidence contours
plotted in the $\gamma$-$\eta$ plane show that discrimination
can only be performed wit 1-1.5$\sigma$ significance.

\end{enumerate}




%
\begin{table}

\begin{tabular}{|>{\centering}p{8cm}|>{\centering}m{2cm}|>{\centering}p{2cm}|>{\centering}p{2cm}|}
\hline 
 & $\sigma_{w_{0}}$ & $\sigma_{w_{1}}$ & FOM\tabularnewline
\hline
\hline 
$\gamma_{0},\gamma_{1}$ fixed & 0.04 & 0.24 & 166\tabularnewline
\hline 
$\Omega_{k}$ fixed and marginalization over both $\gamma_{0},\gamma_{1}$ & 0.12 & 0.36 & 97.3\tabularnewline
\hline 
marginalization over all other parameters & 0.12 & 0.43 & 41.3\tabularnewline
\hline
\end{tabular}\caption{\label{tab:w0_w1}
1  $\sigma$ marginalized errors for the parameters $w_{0}$
and $w_{1}$, obtained with three different methods (reference case, see Fig.~\ref{fig:w0_w1} \label{tab:sigma_w0_w1}).}
\end{table}

%



%

\section{Conclusions}\label{sec:conclusions}
In this paper we addressed the problem of determining the growth rate
of density fluctuations from the estimate of the galaxy power spectrum at
different epochs in future redshift survey. As a reference case we have
considered the proposed EUCLID spectroscopic survey modeled according to the latest,
publicly available survey characteristics~\cite{euclid,geach10}. In this work we
focused on a few issues that we regard as very relevant and that were not
treated in previous, analogous Fisher Matrix analysis mainly aimed at
optimizing the survey setup and the observational strategy. These issues are: {\it i})
the ability in measuring self-consistently galaxy bias with no external information
and the impact of treating the bias as an extra free parameter on the error budget; {\it ii}) the impact of choosing a particular parameterization in determining the growth rate and
in distinguishing dark energy models with very different
physical origins (in particular we focus on the $\Lambda$CDM, $f(R)$ and the DGP, models that are still degenerate with respect to present growth rate data); {\it iii})
the estimate of how errors on the growth rate depend on the degrees of
freedom in the Fisher matrix analysis; {\it iv}) the ability of estimating a possible
coupling between dark matter and dark energy.

The main results of the analysis were already listed in the previous
Section, here we recall the most relevant ones.

\begin{enumerate}
\item With the ``internal bias'' method we were able to estimate bias with 1\% accuracy in a self consistent way using only galaxy positions in redshift-space. The precision in
measuring the bias has a very little dependence on the functional form assumed for $b(z)$. Measuring $b$ with 1\% accuracy will be a remarkable result also from an astrophysical point of view, since it will provide a strong, indirect constraint on the models of galaxy evolution.

\item  We have demonstrated that measuring the amplitude and the slope of the power spectrum in different $z$-bin allows to constrain the growth rate with good accuracy, with no need to assume an external error for $b(z)$. In particular, we found that $s$ can be constrained at $1\sigma$ to within 3\% in each of the 8 redshift bin from $z=0.5$ to $2.1$.
This result is robust to the choice of the biasing function $b(z)$.
The accuracy in the measured $s$ will be good enough  to discriminate among the most popular competing models of dark energy and modified gravity.

\item Taking into account the possibility of a coupling between dark matter and dark energy has the effect of loosening the constraints on the relevant parameters, decreasing the statistical significance in distinguishing models (from $\gtrsim 2\sigma$ to $\lesssim 1.5\sigma$). Yet, this is still a remarkable improvement over the present situation, as can be appreciated from Fig.~\ref{fig:gamma_eta_new_past} where we compare the constraints expected by next generation data to the present ones. Moreover, the {\it Reference} survey will be able to constrain the parameter $\eta$ to within 0.04. Reminding that we can write $\eta=2.1 \beta_c^2$~\cite{diporto08}, this means that the coupling parameter $\beta_c$ between dark energy and dark matter can be constrained to within 0.14, solely employing the growth rate information. This is comparable to existing constraints from the CMB but is complementary since obviously it is obtained at much smaller redshifts. A variable coupling could therefore be detected by comparing the redshift survey results with the CMB ones.
\end{enumerate}

It is worth pointing out that, whenever we have performed statistical tests similar to those already discussed by other authors in the context of a EUCLID-like survey, we did find consistent results.
Examples of this are the values of FOM and errors for $w_0$, $w_1$, similar to those in~\cite{wang10,majerotto11} and the errors on constant $\gamma$ and $w$~\cite{majerotto11}. However, let us notice that all these values strictly depend on the parameterizations adopted and on the numbers of parameters fixed or marginalized over.
In particular, we also found that all these constraints can be improved if one uses additional information from e.g. CMB and other observations. We made a first step in this direction in Fig.~(\ref{fig:histogram}), which shows how the errors on a constant $\gamma$ decrease when progressively more parameters are fixed by external priors.

\section*{Acknowledgements}
This work is supported by the DFG through TRR33 "The Dark Universe". We wish to thank Gigi Guzzo and Andrea Cimatti for insightful suggestions on an earlier draft of this paper.

\clearpage

\providecommand{\href}[2]{#2}\begingroup\raggedright\endgroup


\end{document}